# Cavity Quantum Electrodynamics Design with Single Photon Emitters in Hexagonal Boron Nitride

Yanan Wang[1], Jaesung Lee[1], Jesse Berezovsky[2], and Philip X.-L. Feng[1]*

[1]*Department of Electrical and Computer Engineering, Herbert Wertheim College of Engineering, University of Florida, Gainesville, Florida 32611, USA*

[2]*Department of Physics, Case Western Reserve University, Cleveland, Ohio 44106, USA*

## *Abstract*

Hexagonal boron nitride (h-BN), a prevalent insulating crystal for dielectric and encapsulation layers in two-dimensional (2D) nanoelectronics and a structural material in 2D nanoelectromechanical systems (NEMS), has also rapidly emerged as a promising platform for quantum photonics with the recent discovery of optically active defect centers and associated spin states. Combined with measured emission characteristics, here we propose and numerically investigate the cavity quantum electrodynamics (cavity-QED) scheme incorporating these defect-enabled single photon emitters (SPEs) in h-BN microdisk resonators. The whispering-gallery nature of microdisks can support multiple families of cavity resonances with different radial and azimuthal mode indices simultaneously, overcoming the challenges in coinciding a single point defect with the maximum electric field of an optical mode both spatially and spectrally. The excellent characteristics of h-BN SPEs, including exceptional emission rate, considerably high Debye-Waller factor, and Fourier transform limited linewidth at room temperature, render strong coupling with the ratio of coupling to decay rates $g/\max(\gamma,\kappa)$ predicated as high as 500. This study not only provides insight into the emitter-cavity interaction, but also contributes toward realizing h-BN photonic components, such as low-threshold microcavity lasers and high-purity single photon sources, critical for linear optics quantum computing and quantum networking applications.

---

*Corresponding Author. Email: philip.feng@ufl.edu

Single-photon emitters (SPEs) in solid-state platforms are among the most crucial ingredients for developing integrated quantum photonic circuits,[1,2] which are envisaged to revolutionize information processing, communication, and sensing technologies in the future. Thanks to the advances in materials science and nanotechnology, a proliferation of optically active defect centers have been achieved in wide-bandgap (WBG) materials,[3,4,5,6,7] which can be considered as "inverted atoms" — atomic impurities in otherwise perfect crystals associated with quantized optical transitions. The wide-bandgap attributes encompassing excellent electronic isolation, thermal and chemical stability promise robust single photon emission and exceptional spin coherence even at room temperature.[8,9,10]

Analogous to the SPEs in the conventional WBG materials (*e.g.*, diamond and silicon carbide), the existence of defect-related SPEs in a hallmark van der Waals (vdW) WBG crystal, namely hexagonal boron nitride (h-BN), was reported in late 2015 and sparked growing research interests.[11] Extensive photophysical studies have led to encouraging milestones, including control of single photon emission via electrical/magnetic/strain fields,[12,13,14] optically detected magnetic resonance,[15,16] and Rabi oscillation under resonant excitation.[17] Even though the physical nature of these defect centers is still under investigation, h-BN has been proven as a promising platform to explore light-matter interaction and enable on-demand single photon sources, endowed with large emission rate (>$10^6$ counts/s),[11] strong zero-phonon emission (Debye-Waller factor, $F_{DW}$ ~0.8),[18] high quantum efficiency (~87%),[19] Fourier transform (FT) limited linewidth at room temperature,[18] and single photon purity even at 800 K.[20]

Toward the realization of quantum functionalities based on SPEs, efficient coupling to high-quality optical devices that can direct emission into a single spatial/spectral mode and enhance the emission rate with unit efficiency is requisite. Initial experiments have demonstrated coupling of h-BN defect centers to linear photonic crystal cavities,[21,22] silicon nitride ($Si_3N_4$) microdisk resonators,[23] and dielectric Bragg microcavities,[24] while all in weak coupling regime with Purcell factors ($F_P$) below 10. This is because in the emitter-cavity systems relying on heterogeneous integration,[22,23] the emission dipole is coupled to the evanescent field but not the maximum field confined inside the optical cavity. Moreover, a wide spectral range of emission has been observed from h-BN, with zero-phonon lines (ZPLs) spanning from ultraviolet (UV) to near-infrared (NIR) wavelengths.[25,26,27] Most current cavity designs work at a specific wavelength, and the ones comprising photonic crystal lattices[21,22] and fixed Fabry-Pérot cavities[24] are incompetent to afford tunability over a broad wavelength range. It can be seen that important challenges lie in matching the frequency of cavity resonance with the ZPL and aligning the SPE dipole with the maximum electric field of the optical mode simultaneously.

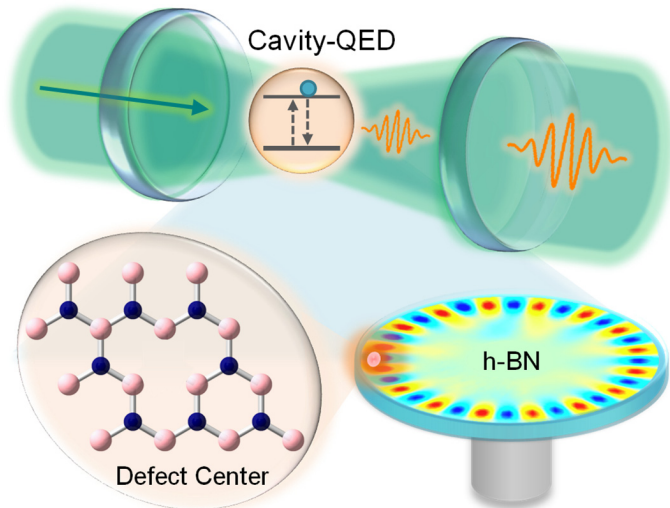

**FIG. 1.** Schematic illustration of a coupled emitter-cavity system consisting of h-BN microdisk cavity and embedded defect center, which can serve as an intriguing platform for exploring the quantum light-matter interactions and facilitating functional quantum photonic devices.

In this Letter, we propose a coupled emitter-cavity system built upon defect center embedded within h-BN microdisk (Fig. 1). The difficulty of spatially and spectrally aligning the SPE with the cavity mode can be resolved in such a whispering-gallery geometry, of which cavity resonance can be flexibly engineered by tuning the radial and azimuthal mode indices. Critical analysis of the ZPL energies of h-BN SPEs and the resonance wavelength, quality ($Q$) factor, mode volume as functions of microdisk geometry have been performed. We further relate these parameters to the coupling and decay rates in cavity quantum electrodynamics (cavity-QED), leading to a quantitative and comprehensive understanding of the emitter-photon interaction in strong coupling regimes.

To address the frequency matching issue raised by the broad range of emission, we start with surveying the spectral distribution of emission from 560 nm to 750 nm. It is noticeable that we exclude the emitters with high photon energies (4–6 eV) in this study, because they may originate from localized exciton transitions close to the h-BN bandgap (~5.9 eV).[28] We evaluate the emission spectra from dozens of h-BN flakes mechanically exfoliated and then transferred onto patterned silicon dioxide on silicon (290nm $SiO_2$ on Si) substrates, by utilizing custom-built confocal microscope systems and spectroscopic techniques.[29] The systems consist of high-magnification optical objectives (50× and 100×), and spectrometers (Horiba iHR550 and Princeton Instruments 2500) with charge-coupled device (CCD) cameras sufficiently sensitive to record weak emission signals. Flakes possessing optically active defect centers produce bright fluorescence that can be directly visualized in fluorescence images with the below-bandgap excitation from a sub-milliwatt 532 nm laser. In our experiments, individual spectra are collected from isolated bright spots in fluorescence images with spatial resolution below 1 μm. Correlation measurements are performed by using the Hanbury Brown and Twiss (HBT) interferometry with single photon counting avalanche photodiodes (Excelitas Technologies Inc.) and a timing module (PicoQuant Inc.), to further verify the single photon nature of the emission process.

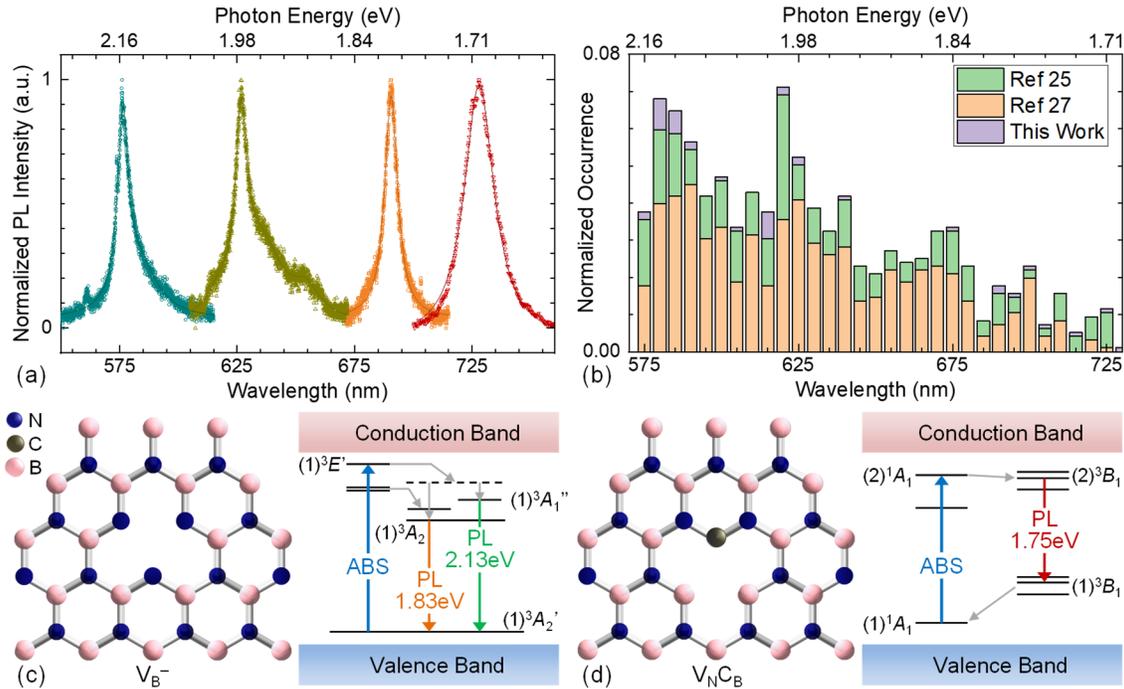

**FIG. 2.** Spectral distribution of emission and proposed defect centers in h-BN. (a) Typical emission spectra of defect-related SPEs in mechanically exfoliated h-BN flakes. The raw spectra (*Hollow symbols*) are processed with baseline subtraction, normalized to unity intensity, and fitted to Lorentz peak functions (*Solid lines*). (b) Histogram plot of ZPL energies for h-BN SPEs based on data in this work (*Purple*), Ref. 25 (*Light green*), and Ref. 27 (*Orange*). The occurrence is normalized by dividing the total number of emitters. Illustrations of (c) Negatively charged boron vacancy $V_B^-$ and (d) Carbon-related center $V_N C_B$, and their corresponding defect states within the bandgap (not in scale, including data from Refs. 31&32). ABS: Absorption; PL: Photoluminescence.

Figure 2(a) displays four typical spectra from our measurements at room temperature, most of which can be fitted into two Lorentz peaks with a relatively sharp line and a broader feature in lower energy, assigned as zero-phonon line (ZPL) and phonon sideband (PSB), respectively. It can be seen that the energies of different emitters spread over a considerably broad spectral range, while the ZPL wavelengths cluster in four distinct positions, with peaks around 575 nm, 625 nm, 680 nm, and 725 nm and linewidth below 10 nm. We summarize in Fig. 2(b) the data from our observations, into a histogram of occurrences, combined with data from measurements at cryogenic temperatures.[25,27] Each cluster of emission could originate from a specific defect composition and involve at least one defect level lying deep within the semiconductor bandgap. However, the variations in local environment such as strain and electric field from trapped charges cause shifts or splitting of defect states and inhomogeneous broadening of emission, accounting for a nearly continuous distribution in ZPL energies.

An in-depth understanding of SPEs could be gained by comparing them with *ab initio* calculations.[30,31,32] Distinct from the well-studied defect-related SPEs in three-dimensional (3D) WBG crystals, such as nitrogen vacancy (NV) centers in diamond,[3] silicon vacancy ($V_{Si}$) and divacancy (DV) centers in silicon carbide,[4,5] the atomic origin of emission in h-BN is still under debate. Here, we consider two types of defect sites, negatively charged boron vacancy ($V_B^-$) and carbon-related center ($V_NC_B$), which have been proposed to interpret the recent observation of optically detected magnetic resonance (ODMR).[15,16,31,32] As predicted in Ref. 31, $V_B^-$ center could give rise to six intrinsic orbital levels and the triplet transition $(1)^3E' \rightarrow (1)^3A_2'$ with dominant $(1)^3A_2 \rightarrow (1)^3A_2'$ component, of which transition energy $\Delta E$=1.78–1.83 eV agrees well with the emission detected around 680 nm (Fig. 2(c)). Similarly, the transition $(1)^3A_1'' \rightarrow (1)^3A_2'$ is calculated to possess an energy $\Delta E$=2.00−2.13 eV, corresponding to the ZPLs between 580 nm and 620 nm. Spin-triplet $(2)^3B_1$ to $(1)^3B_1$ transition of the $V_NC_B$ center, consisting of one nitrogen vacancy and one neighboring carbon atom substitution,[32] associates with ZPL energy of ~1.75 eV and wavelength at ~710 nm (Fig. 2(d)).

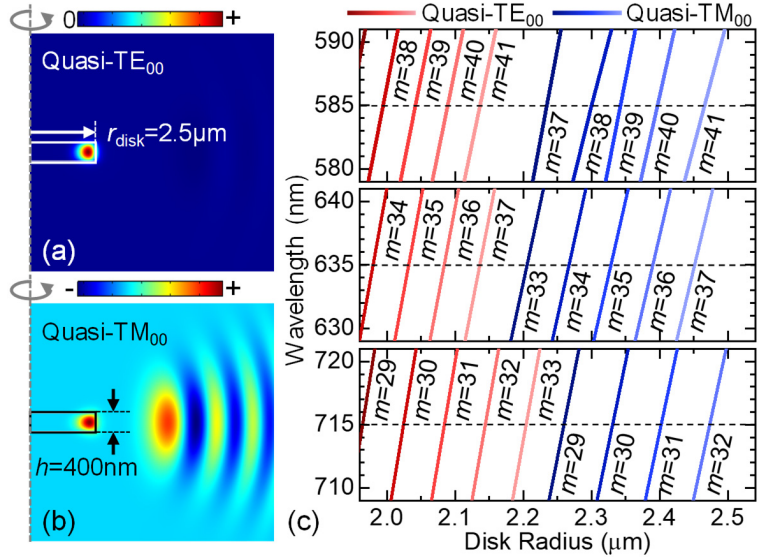

**FIG. 3.** Whispering-gallery modes of h-BN microdisk cavity at different ZPL wavelengths. Electric-field distribution of (a) Quasi-TE$_{00}$ and (b) Quasi-TM$_{00}$ modes for a microdisk ($r_{disk}$=2.5 µm, $h_{disk}$=400 nm) around 585 nm. (c) FEM-simulated cavity resonances as functions of disk radius and azimuthal mode number around 585 nm, 635 nm, 715 nm, respectively. *Red solid lines*: Quasi-TE$_{00}$ mode; *Blue solid lines*: Quasi-TM$_{00}$ mode.

In accordance with the spectral distribution of ZPLs discussed above, we explore the resonance behaviors of microdisk at 585 nm, 635 nm, and 715 nm, via finite-element eigenfrequency simulations. Taking advantage of the azimuthal symmetry, only a two-dimensional cross-section of microdisk is simulated by employing the 2D axisymmetric space dimension in COMSOL Multiphysics$^{TM}$.[33] It is worthwhile noting that the wavelength-dependent refractive indices and birefringence of h-BN are applied in our numerical simulations. As demonstrated in a recent report,[34] h-BN flakes exhibit significant birefringence due to the peculiar vdW layered structure, and in-plane refractive index $n_\parallel$ and out-of-plane index $n_\perp$ can be calculated using Sellmeier's equations:

$$n_\parallel^2(\lambda)=1+\frac{3.336\lambda^2}{\lambda^2-26322}, \quad n_\perp^2(\lambda)=1+\frac{2.2631\lambda^2}{\lambda^2-26981}, \tag{1}$$

where the wavelength λ is in nanometers. Relatively small out-of-plane refractive index (~1.8) compared to the in-plane index (~2.1) over the wavelengths of interest results in comparatively weak confinement of modes along the direction perpendicular to the h-BN layers, which can be directly visualized in the electric-field distribution as shown in Fig. 3(a) & 3(b).

Whispering-gallery modes of microdisk resonator follow the relation $m\lambda=2\pi r_{disk}n_{eff}$, where $m$ is the azimuthal index, $\lambda$ is the resonance wavelength, $r_{disk}$ is the microdisk radius, $n_{eff}$ is the effective refractive index. For device design, the resonance wavelength can be precisely engineered by tuning the disk radius and azimuthal index. Figure 3(c) depicts the capability of matching the fundamental quasi-transverse-electric (quasi-TE$_{00}$) and quasi-transverse-magnetic (quasi-TM$_{00}$) modes with different ZPL wavelengths at the same time. For instance, a microdisk with radius $r_{disk}$=2.1 μm and thickness $h_{disk}$=400 nm can support quasi-TE$_{00}$ mode around 585 nm, 635 nm, and 715 nm with azimuthal index $m$=40, 36, and 31, respectively. The free spectral range (FSR) of such a microdisk at 635 nm is calculated as 14.3 nm. Although the predicated FSR is smaller than the wavelength difference between ZPL and PSB reported in Ref. 18, the large Debye-Waller factor of 0.8 and high quantum efficiency of ~87% can ensure that the coupling between the cavity mode and ZPL is dominant. The PSB can be further filtered out by designing the microdisk-waveguide or microdisk-fiber coupling schemes in real measurements.

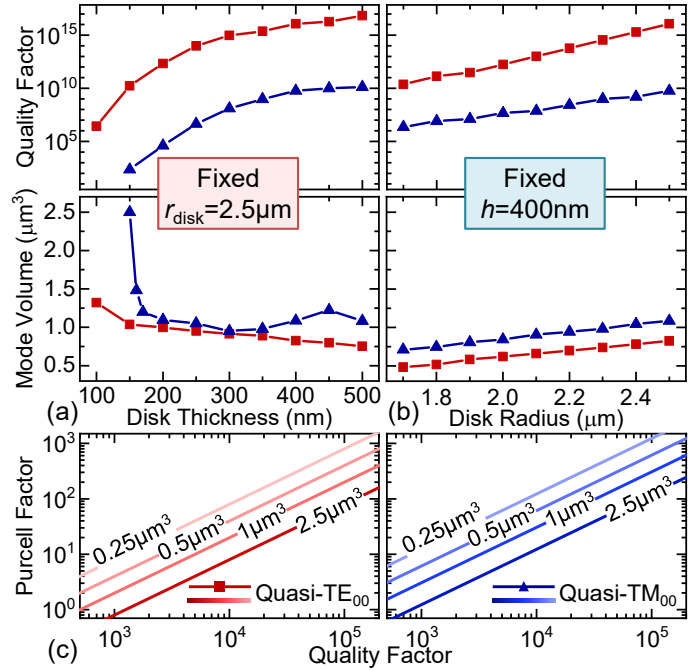

**FIG. 4.** Geometry dependence of mode confinement in h-BN microdisk cavity. Radiation-limited quality factor $Q_{rad}$ and effective mode volume $V_{eff}$ as functions of microdisk (a) thickness and (b) radius around 635 nm. (c) Purcell factor scaling as a function of $Q$ and $V_{eff}$ around 635 nm. *Red solid lines & Red squares*: Quasi-TE$_{00}$ mode; *Blue solid lines & Blue triangles*: Quasi-TM$_{00}$ mode.

Further characterization of microdisk cavity is focused on resonances around 635 nm, at which FT limited lines from h-BN SPEs have been reported under resonant excitation at room temperature.[18] To determine the cut-off thickness of microdisk, the mechanical properties of h-BN are also taken into consideration. A minimum thickness of 100 nm is set based on our previous studies to ensure that the h-BN flake is rigid enough to facilitate a pedestal-supported structure.[35,36] From the mode analysis performed with a fixed microdisk radius $r_{disk}$=2.5 µm, we can notice that the radiation-limited quality factor ($Q_{rad}$) merely reaches $10^2$ for quasi-TM$_{00}$ mode at ~635 nm, when the microdisk thickness $h_{disk}$ is below 150 nm (Fig. 4(a)). As $h_{disk}$ increases to 500 nm, $Q_{rad}$ escalates and saturates around $10^{16}$ and $10^9$ for quasi-TE$_{00}$ and quasi-TM$_{00}$ modes, respectively. Significantly lower $Q_{rad}$ of quasi-TM$_{00}$ mode confirms weak confinement resulting from smaller refractive index along out-of-plane direction. Effective mode volume ($V_{eff}$) below µm$^3$ can be achieved for microdisk with $h_{disk}$>150 nm. In parallel, the influence of microdisk radius $r_{disk}$ is investigated with thickness $h_{disk}$ fixed at 400 nm (Fig. 4(b)). Both quasi-TE$_{00}$ and quasi-TM$_{00}$ modes exhibit exponential dependence of $Q_{rad}$ and approximately linear dependence of $V_{eff}$ on $r_{disk}$ in the study range. For quasi-TE$_{00}$ mode, $V_{eff}$ can be as small as 0.5 µm$^3$ (equal to ~2$\lambda^3$), while maintaining $Q_{rad}$ over $10^{10}$, at which level radiation losses are not expected to be the dominant loss mechanism. To date, the measured values from the h-BN microdisk with similar dimensions are in order of $Q$ ~$10^3$,[37,38] presumably limited by the fabrication imperfections.

In a coupled emitter-cavity system, the spatial mode density inside the cavity can be altered substantially; hence the spontaneous emission rate can be either enhanced ($F_P$>1) or inhibited ($F_P$<1), determined by the Purcell factor

$$F_p \equiv \frac{3}{4\pi^2}\left(\frac{\lambda}{n_0}\right)^3 \frac{Q}{V_{eff}}, \qquad (2)$$

where $n_0$ is the refractive index at the location of the emitter. Purcell factor as a function of quality factor $Q$ and effective mode volume $V_{eff}$ is plotted as Fig. 4(c). With sub-µm$^3$ mode volume, enhancement ($F_P$>1) can be realized even when $Q$=$10^3$, and $F_P$ can be as large as 300 if $Q$ is improved to $10^5$. Since the probability of spontaneous emission placing a photon into the cavity is given by $\beta$=$F_P$/($F_P$+1), deterministic photon emission into a single field mode is possible, if $F_P$ is sizable to make $\beta$≈1. The potential coupling strength can be described by the cooperativity parameter

$$C \equiv \frac{3}{4\pi^2}\left(\frac{\lambda}{n_0}\right)^3 \frac{Q}{V_{eff}} \frac{\gamma_{zpl}}{\gamma_{total}} \left|\frac{\vec{E}(\vec{r}_d)}{\vec{E}(\vec{r}_m)}\right|^2 = F_p \frac{\gamma_{zpl}}{\gamma_{total}} \left|\frac{\vec{E}(\vec{r}_d)}{\vec{E}(\vec{r}_m)}\right|^2, \qquad (3)$$

where $\gamma_{zpl}$ is the rate of zero-phonon emission at wavelength $\lambda$ and $\gamma_{total}$ is the total rate of spontaneous emission, $\gamma_{zpl}/\gamma_{total}$=$F_{DW}$. $\left|\frac{\vec{E}(\vec{r}_d)}{\vec{E}(\vec{r}_m)}\right|^2$ represents the overlapping of the emitter dipole and the maximum electric field inside the cavity. Applying the reported Debye-Waller factor $F_{DW}$ ~0.8 and assuming a perfect alignment between the emitter and cavity mode, a theoretical expectation of the cooperativity parameter is over 200 for a microdisk with $Q$=$10^5$.

It is known that $Q$ and $V_{eff}$ describe the cavity decay rate ($\kappa$) and peak electric field strength within the cavity, respectively. To further translate the cavity mode analysis performed above to the standard parameters studied in cavity-QED, we evaluate the cavity decay rate $\kappa/2\pi$=$\omega/4\pi Q$ and coherent coupling rate

$$g/2\pi = \frac{1}{2\tau_{sp}}\sqrt{\frac{3c\lambda^2\tau_{sp}}{2\pi n_0^3 V_{eff}}}, \qquad (4)$$

where $\tau_{sp}$=2.65 ns is spontaneous emission lifetime and the value is adopted from the FT limited lines with emitter dephasing rate $\gamma/2\pi$=63 MHz.[18] We also assume $Q$ = $Q_{rad}$, and the defect center is optimally positioned within the cavity field. As shown in Fig. 5(a) & 5(b), The cavity decay rate decreases as the thickness and radius of the microdisk increase for both quasi-TE$_{00}$ and quasi-TM$_{00}$ modes. The coupling gradually strengthens with increasing thickness, while becomes weaker with enlarging radius.

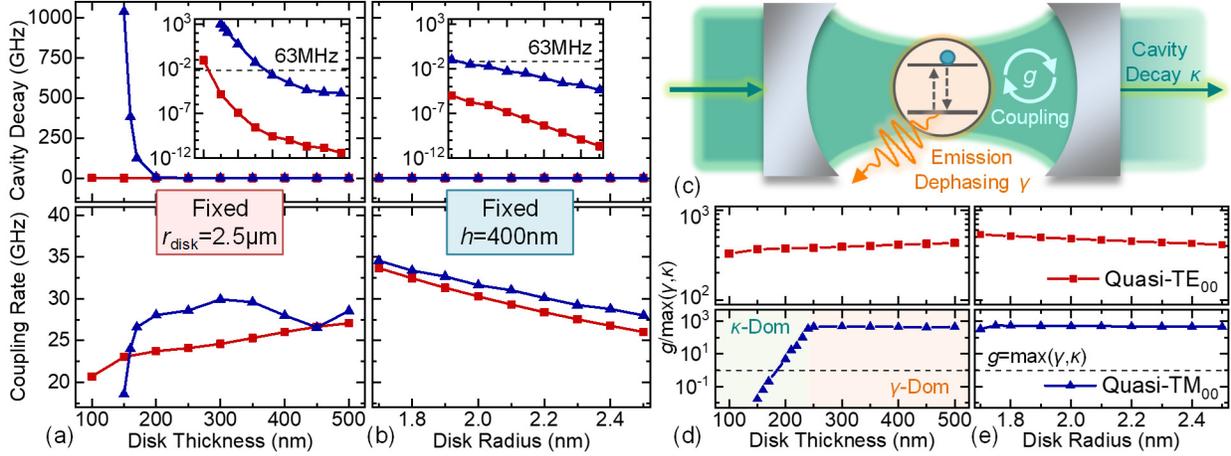

**FIG. 5.** Cavity-QED analysis. Cavity decay rate $\kappa/2\pi$ and coherent coupling rate $g/2\pi$ as functions of microdisk (a) thickness and (b) radius. *Insets*: Cavity decay rate in semi-log plots with *y*-axis unit of GHz. *Black dashed lines*: Emitter dephasing rate $\gamma/2\pi$=63 MHz. (c) Schematic of the cavity-QED and the interaction between an optical cavity and a two-level system. Plots of the ratio between the calculated coupling rate *g* and the maximum decay rate of the defect-cavity system max($\gamma,\kappa$), as functions of microdisk (d) thickness and (e) radius. *Black dashed lines*: $g$=max($\kappa,\gamma$).

By comparing the coupling rate $g/2\pi$ to the cavity decay rate $\kappa/2\pi$ and the emitter dephasing rate $\gamma/2\pi$, we can further determine whether a system is in strong coupling or not (Fig. 5(c)). It is noticeable that for the proposed system, strong coupling regime with $g$>max($\kappa,\gamma$) can be reached as thickness $h_{disk}$ exceeds 200 nm (Fig. 5(d) & 5(e)). Because dissipation in a strongly coupled emitter-cavity system either results from cavity decay or emitter dephasing, the ratio of $g$ to the maximum decay rate in the system $g/\max(\kappa,\gamma)$ is calculated, representing the number of coherent exchanges of energy (Rabi oscillations) between the emitter and cavity. Except for the microdisk with $h_{disk}$<250 nm (or $r_{disk}$<1.7 μm), the loss is dominated by emitter dephasing and $g/\max(\kappa,\gamma)$ plateaus at a value above 500.

In summary, we have synergized the experimental and analytical efforts to understand the defect-related single photon emission in h-BN. Utilizing the spectral configurability benefitted from whispering-gallery geometry, we have proposed a coupled emitter-cavity system based on h-BN SPE embedded within h-BN microdisk, numerically explored the cavity characteristics, and provided the design and theoretical evaluation of light-matter interaction via cavity-QED analysis. The methodology and quantitative metrics developed in this study can serve as practical guidelines for designing and facilitating h-BN photonic devices and integrated systems toward quantum science and engineering applications.

### Acknowledgments

We are thankful for the support from the National Science Foundation (NSF) via EFRI ACQUIRE program (Grant EFMA 1641099), its Supplemental Funding through the Research Experience and Mentoring (REM) program.